\documentclass[12pt,a4paper]{article}
\usepackage{epsfig}
\usepackage{amsmath,amssymb}

\def\nostrocostrutto#1\over#2{\mathrel{\mathop{\kern 0pt \rlap 
  {\raise.2ex\hbox{$#1$}}}
  \lower.9ex\hbox{\kern-.190em $#2$}}}

\def\GeV{{\rm GeV}}
\def\MeV{{\rm MeV}}
\def\eV{{\rm eV}}
\def\ra{ \rightarrow }
\def\qq{{q\bar{q}}}
\def\bb{{b\bar{b}}}

\catcode`@=11
\newcount\@tempcntc
\def\@citex[#1]#2{\if@filesw\immediate\write\@auxout{\string\citation{#2}}\fi
  \@tempcnta\z@\@tempcntb\m@ne\def\@citea{}\@cite{\@for\@citeb:=#2\do
    {\@ifundefined
       {b@\@citeb}{\@citeo\@tempcntb\m@ne\@citea\def\@citea{,}{\bf ?}\@warning
       {Citation `\@citeb' on page \thepage \space undefined}}%
    {\setbox\z@\hbox{\global\@tempcntc0\csname b@\@citeb\endcsname\relax}%
     \ifnum\@tempcntc=\z@ \@citeo\@tempcntb\m@ne
       \@citea\def\@citea{,}\hbox{\csname b@\@citeb\endcsname}%
     \else
      \advance\@tempcntb\@ne
      \ifnum\@tempcntb=\@tempcntc
      \else\advance\@tempcntb\m@ne\@citeo
      \@tempcnta\@tempcntc\@tempcntb\@tempcntc\fi\fi}}\@citeo}{#1}}
\def\@citeo{\ifnum\@tempcnta>\@tempcntb\else\@citea\def\@citea{,}%
  \ifnum\@tempcnta=\@tempcntb\the\@tempcnta\else
   {\advance\@tempcnta\@ne\ifnum\@tempcnta=\@tempcntb \else \def\@citea{--}\fi
    \advance\@tempcnta\m@ne\the\@tempcnta\@citea\the\@tempcntb}\fi\fi}
\catcode`@=12

\begin{document}

\setcounter{page}{1}
\thispagestyle{empty}

\def\jet{{\mbox{\scriptsize jet}}}
\def\cO#1{{\cal O}\left(#1\right)}
\def\lrang#1{\left\langle#1\right\rangle}
\def\br{brems\-strah\-lung}
\def\Br{Brems\-strah\-lung}
\def\QQ{\relax\ifmmode{Q\overline{Q}}\else{$Q\overline{Q}${ }}\fi}
\def\qq{\relax\ifmmode{q\bar q}\else{$q\bar q${ }}\fi}
\def\abs#1{\left|#1\right|}

\def\la{\mathrel{\mathpalette\fun <}}
\def\ga{\mathrel{\mathpalette\fun >}}
\def\fun#1#2{\lower3.6pt\vbox{\baselineskip0pt\lineskip.9pt
  \ialign{$\mathsurround=0pt#1\hfil##\hfil$\crcr#2\crcr\sim\crcr}}}

\def\eV{{\rm e\kern-0.12em V}}
\def\MeV{{\rm M}\eV} \def\keV{{\rm k}\eV}
\def\GeV{{\rm G}\eV} \def\TeV{{\rm T}\eV}

\newcommand\beeq{\begin{eqnarray}}
\newcommand\eeeq{\end{eqnarray}}
\def\as{\alpha_s}

\begin{flushright}
{IPPP/05/36 \\
DCPT/05/72 \\
4th July 2005\\}

\end{flushright}

\vspace*{0.5cm}

\begin{center}
{\Large \bf Diffractive processes as a tool for searching for new physics\footnote{To be published in
the Proc. of the Gribov-75 Memorial Workshop, Budapest,  May 2005.}}

\vspace*{1cm}
\textsc{V.A.~Khoze$^{a,b}$, A.B.~Kaidalov$^{a,c}$, A.D.~Martin$^a$, M.G.~Ryskin$^{a,b}$ and W.J.~Stirling$^a$} \\

\vspace*{0.5cm}
$^a$ Institute for
Particle Physics Phenomenology, \\
University of Durham, DH1 3LE, UK \\
$^b$ Petersburg Nuclear Physics Institute, Gatchina,
St.~Petersburg, 188300, Russia \\
$^c$ Institute of Theoretical and Experimental Physics, Moscow, 117259, Russia\\

\end{center}

\vspace*{0.5cm}

{\bf Abstract:} We show that the addition of detectors to tag the outgoing forward protons, at the LHC, will significantly
enlarge the potential of studying New Physics.
A topical example is
Higgs production by the exclusive double-diffractive process, $pp \to p+H+p$. We discuss the production of Higgs bosons in both the SM and MSSM. We show how the predicted rates may be checked at the Tevatron by observing the exclusive double-diffractive production of dijets, or $\chi_c$ or $\chi_b$ mesons, or $\gamma \gamma$ pairs.


 


\section{Introduction}

The use of forward proton detectors as a means to study Standard Model (SM) and New Physics
at the LHC has only been fully appreciated within the last few years; see, for
example \cite{INC,cox1,kp,prs,mb} and references therein.
By detecting protons that have lost less than about 2\%  of their longitudinal
momentum, a rich QCD, electroweak, Higgs and BSM programme becomes accessible, with
a potential to study phenomena which are unique to the LHC, and difficult even at a future
linear collider \cite{fp420}.

In particular, the so-called central exclusive production (CEP) processes
may provide a 
very friendly environment to search for, and identify 
the nature of, new particles at the LHC; in particular, Higgs bosons.
There is also a potentially rich, more exotic, physics menu 
including (light) gluino and squark production,
 gluinonia, radions, and indeed any object which has $0^{++}$  (or $2^{++}$) quantum 
numbers and couples strongly to gluons \cite{INC}. 
By central exclusive, we mean the process $pp\rightarrow p + X + p$, where 
the + signs denote the absence of hadronic activity (that is, the presence of a rapidity gap) between the outgoing protons and the 
decay products of the central system $X$.

It is a pleasure to recall that the whole strategy of predicting
diffractive phenomena, and, in particular, of 
CEP  processes, is based on the ideas developed by V.N. Gribov.
 We list only some of these: Regge poles in particle physics, 
 the vacuum pole (Pomeron) and its shrinkage, Glauber-Gribov theory of multiple scattering,
 Gribov's reggeon calculus, Gribov's factorization,
 the Abramovsky-Gribov-Kancheli cutting rules, Gribov's theorem for
 bremsstrahlung at high energies, the Gribov-Lipatov (DGLAP) evolution
 equations, the Frolov-Gorshkov-Gribov-Lipatov approach to Regge processes in gauge theories,
 and much, much more.

There are three main reasons why CEP is especially attractive for searches for
new heavy objects.
First, if the outgoing protons remain intact and scatter through small angles then, to a very good approximation,
the primary active di-gluon system obeys a $J_z=0$, C-even, P-even, selection rule
\cite {Liverpool,KMRmm}. Here $J_z$ is the projection of the total angular momentum
along the proton beam axis. This selection rule readily permits a clean determination 
of the quantum numbers of the observed new
(for example, Higgs-like) resonance, which will be dominantly produced in a scalar state.
Secondly, because the process is exclusive, the energy loss of the outgoing protons is directly
related to the mass of the central system, allowing a potentially excellent mass resolution, irrespective 
of the decay mode of the produced particle.\footnote{Recent studies suggest that the missing mass resolution 
$\sigma$ will be of order $1\%$ for a $140$~GeV 
central system, assuming both the outgoing protons are detected at 420m from the interaction point \cite{cox1,fp420}.}
Thirdly, a signal-to-background ratio of order 1
(or even better) is achievable 
\cite{DKMOR,cox1}.
This ratio becomes significantly larger for the 
lightest Higgs boson in certain regions of the MSSM parameter space 
\cite{KKMR04}. 

Moreover, in some MSSM Higgs scenarios CEP provides
an opportunity for a lineshape analysis \cite{KKMR04,je}.
Another attractive feature is the ability to directly 
probe the CP-structure of the  Higgs sector by measuring the azimuthal asymmetry 
of the outgoing tagged protons 
\cite{Khoze:2004rc}. A different strategy, to explore the manifestation of
explicit CP-violation in the Higgs sector, was recently studied by Ellis et al.
\cite{je}.

It is worth mentioning that,
by tagging both of the outgoing protons, the LHC is effectively turned into a gluon-gluon collider.
This will open up a rich, `high-rate' QCD physics menu (especially
concerning diffractive phenomena), which will allow the study of the skewed, unintegrated
gluon densities, as well as the details of rapidity gap survival; see, for example, \cite{INC,kmrBH}. 
Note that CEP provides a source of practically
pure gluon jets; that is we effectively have a `gluon factory' \cite{KMRmm}. This can be an ideal laboratory in which
to study the properties of gluon jets, especially in comparison with the
quark jets, and will even allow a search for glueballs.
The forward-proton-tagging  approach also offers a unique programme of
high-energy photon-interaction physics at the LHC; see, for example, \cite{kp,KMRphot}.

The `benchmark' CEP process, for these new physics
searches, is Higgs production. In the mass range
around 115-130 GeV, its detection
 at the LHC will not be an easy task. 
There is no obvious perfect detection
process, but rather a range of possibilities, none of which is
compelling on its own.  {\em Either}
large signals are accompanied by a huge background, {\em or} the
processes have comparable signal and background rates for which
the number of Higgs events is rather small.
The predicted cross section for the CEP production of 
a SM Higgs, with mass 120 GeV, at the LHC is 
3 fb, falling to 1 fb for a mass of 200 GeV; see \cite{KMR}.

 From an experimental perspective, the $WW$ decay channel is the simplest way to observe the SM Higgs in the tagged-proton approach \cite{krs,cox2}.
The  $b \bar b$ decay channel is more challenging from a trigger perspective, although, in this case, the `useful' event rate is more favourable for masses below about 130 GeV.
Moreover, the latter decay mode becomes extremely important in the so-called
intense coupling regime \cite{bdn} of the MSSM, where CEP is
likely to be the discovery channel \cite{KKMR04}. In this case, we expect about 10$^3$
exclusively produced double-tagged Higgs bosons for 30 fb$^{-1}$ of delivered luminosity.
About 100 would survive the experimental cuts \cite{DKMOR}, with a signal-to-background
ratio of the order of 10. 

In the case of the exclusive process, $pp\ra p + H + p$,
a major experimental task is to provide
a set-up in which the bulk of the proton-tagged Higgs signal is deposited in a
smallest possible missing-mass window; $\Delta M_{\rm missing}$ of about 3 GeV should be achievable. 
Note that for the $b \bar b$ channel the CEP process allows the mass of
the Higgs to be measured in two independent ways. First, the tagged
protons give $M_H = M_{\rm missing}$ and second, via the $H\ra\bb$
decay, we have $M_H = M_{\bb}$, although now the resolution is
much poorer, with $\Delta M_{\bb}\simeq10$~GeV or more. The existence of
matching peaks, centered about $M_{\rm missing}=M_{\bb}$, is a
unique feature of the exclusive diffractive Higgs signal. Besides
its obvious value in identifying the Higgs, the mass equality also
plays a key role in reducing background contributions. Another crucial
advantage of the exclusive process $pp\ra p+H+p$, with $H\ra\bb$,
is that the leading order $gg\ra\bb$ background subprocess is
suppressed by the $J_z=0$, P-even selection rule \cite{KMRmm,DKMOR}.

\section{Calculation of the exclusive Higgs signal}

\begin{figure}
\begin{center}
\centerline{\epsfxsize=0.4\textwidth\epsfbox{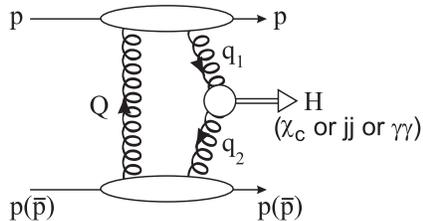}}
\caption{Schematic diagram for central exclusive  production,
$pp \to p+X+p$. The presence of Sudakov form factors ensures the infrared
stability of the $Q_t$ integral over the gluon loop. It is also necessary
to compute the probability, $S^2$, that the rapidity gaps survive soft rescattering.}
\label{fig:H}
\end{center}
\end{figure}

The basic mechanism for the exclusive process, $pp\ra p+H+p$, is
shown in Fig.~$\ref{fig:H}$. Since the dominant contribution comes from the
region $\Lambda_{\rm QCD}^2\ll Q_t^2\ll M_H^2$, the amplitude may
be calculated using perturbative QCD techniques \cite{KMR,KMRmm}
\begin{equation}
{\cal M}_H \simeq N\int\frac{dQ^2_t\ V_H}{Q^6_t}\: f_g(x_1, x_1', Q_t^2, \mu^2)f_g(x_2,x_2',Q_t^2,\mu^2), \label{eq:M}
\end{equation}
where the overall normalization constant $N$ can be 
written in terms of the $H\to gg$ decay width \cite{INC,KMR}, and where the
$gg\to H$ vertex factors for CP $=\pm 1$ Higgs production are, after azimuthal-averaging,
\begin{equation}
V_{H(0^+)} \simeq  Q_t^2, ~~{\rm and}~~
V_{A(0^-)}  \simeq  (\vec p_{1t} \times \vec p_{2t})\cdot \vec n_0,
\label{eq:rat4}
\end{equation}
Expressions (\ref{eq:M},\ref{eq:rat4}) hold for small $p_{it}$, where the $\vec p_{it}$
are the transverse momenta of the outgoing protons,
and $\vec n_0$ is a unit vector in the beam direction.
The $f_g$'s are the skewed unintegrated gluon densities at the hard scale $\mu$,
taken to be $M_H/2$. Since $(x'\sim Q_t/\sqrt s)\ll (x\sim M_H/\sqrt s)\ll 1$, it is possible
to express 
$f_g(x,x',Q_t^2,\mu^2)$, to single log accuracy, in
terms of the conventional integrated density $g(x)$. The $f_g$'s
embody a Sudakov suppression factor $T$, which ensures that the gluon does not radiate in the
evolution from $Q_t$ up to the hard scale $M_H/2$, and
so preserves the rapidity gaps.  The apparent infrared divergence of~(\ref{eq:M}) is nullified
 for $H(0^+)$ production by these Sudakov factors.\footnote{Note also that the
Sudakov factor inside the loop integration induces an additional strong
 decrease (roughly as $M^{-3}$ \cite{KKMR04}) of the cross section as 
the mass $M$ of the centrally produced hard system increases.
Therefore, the price to pay for neglecting this suppression effect
would be to considerably overestimate the CEP cross section at large masses.}
However, the amplitude for $A(0^-)$ production is much more sensitive to the infrared contribution. Indeed,
let us consider the case of small $p_{it}$ of the outgoing protons. Then we see, from~(\ref{eq:rat4}), that
 the $dQ^2_t/Q_t^4$ integration for $H(0^+)$ is
replaced by $p_{1t}p_{2t} dQ^2_t/Q_t^6$ for $A(0^-)$, and now the Sudakov suppression is not
enough to prevent a significant contribution from the low $Q_t^2$ domain.

The radiation associated with the $gg\ra H$ hard
subprocess is not the only way to populate and to destroy the
rapidity gaps. There is also the possibility of soft rescattering
in which particles from the underlying event populate the gaps.
The probability, $S^2=0.03$ at the LHC, that the gaps survive the soft
rescattering was calculated using a two-channel eikonal model,
which incorporates high mass diffraction\cite{KMRsoft}. Including
this factor, and the NLO $K$ factor, the cross section is
predicted to be \cite{INC,KMR}
\begin{equation}
\sigma(pp\ra p+H+p)\simeq 3\:{\rm fb} \label{eq:sigma}
\end{equation}
for the production of a SM Higgs boson of mass 120~GeV
at the LHC. We evaluated
that there may be a factor of 2.5 uncertainty (up or down) in this
prediction\cite{KKMR04}.

If we include a factor 0.6 for
the efficiency associated with proton tagging, 0.67 for the $H\ra\bb$ branching fraction,
0.6 for $b$ and
$\bar{b}$ tagging, 0.5 for the $b,\bar{b}$ jet polar angle cut,
$60^\circ<\theta<120^\circ$, (necessary to reduce the $\bb$ QCD
background)\cite{DKMOR}, then,
for a luminosity of ${\cal L}=30\:{\rm fb}^{-1}$, the original $3 \times 30=90$
 events are reduced to an observable
signal of 11 events.

\section{Background to the exclusive $H\ra\bb$ signal}

The advantage of the $p+(H\ra\bb)+p\,$ signal is that there exists
a $J_z=0$ selection rule, which requires the leading order
$gg^{PP}\ra\bb$ background subprocess to vanish in the limit of
massless quarks and forward outgoing protons. (The $PP$ superscript is
to note that each gluon comes from colour-singlet $gg~t$-channel exchange.) However,
in practice, LO background contributions remain. The prolific
$gg^{PP}\ra gg$ subprocess may mimic $\bb$ production since we may
misidentify the outgoing gluons as $b$ and $\bar{b}$ jets.
Assuming the expected 1\% probability of misidentification, and
applying $60^\circ<\theta<120^\circ$ jet cut, gives a
background-to-signal ratio $B/S \sim 0.18$. (Here, for reference, we
assume that the mass window over which we collect the signal, $\Delta M\sim 3\sigma=3$ GeV).

Secondly, there is an
admixture of $|J_z|=2$ production, arising from non-forward going
protons which gives $B/S \sim 0.24$. Thirdly, for a massive quark
there is a contribution to the $J_z=0$ cross section of order
$m_b^2/E_T^2$, leading to $B/S \sim 0.18$, where $E_T$ is the
transverse energy of the $b$ and $\bar{b}$ jets.\footnote{There are reasons
to hope that, due to higher-order QCD effects, this particular background contribution
will be a few times smaller.}

Next, we have the possibility of NLO $gg^{PP}\ra\bb g$ background
contributions, which for large angle, hard gluon radiation does not obey the selection rules. Of course, the extra gluon may be observed
experimentally and these background events eliminated. However,
there are exceptions. The extra gluon may go unobserved in the
direction of a forward proton. This background may be effectively
eliminated by requiring the equality $M_{\rm missing} = M_{\bb}$.
Moreover, soft gluon emissions from the initial $gg^{PP}$ state
factorize and, due to the overriding $J_z=0$ selection rule, these
contributions to the QCD
$\bb$ production are also suppressed. The remaining danger is
large angle hard gluon emission which is collinear with either the
$b$ or $\bar{b}$ jet, and, therefore, unobservable. If the cone
angle needed to separate the $g$ jet from the $b$ (or $\bar{b}$)
jet is $\Delta R \sim 0.5$, then the expected background from
unresolved three-jet events leads to $B/S \simeq 0.18$.
The NNLO $\bb gg$ background contributions are found to be
negligible (after requiring $M_{\rm missing}\simeq M_{\bb}$), as
are soft Pomeron-Pomeron fusion contributions to the background
(and to the signal)~\cite{DKMOR}.  Also note that radiation off the
screening gluon, in Fig.~$\ref{fig:H}$, is numerically small \cite{myths}.

\section{The signal-to-background ratio for $H\ra\bb$ mode}

So, in total, for the exclusive production of a 120 GeV (SM) Higgs boson at the LHC 
with the integrated luminosity ${\cal L}=30$ fb$^{-1}$, after cuts and acceptances
we can expect about 10 events, with a
signal-to-background
ratio $S/B$ of the order of 1.
In the case of a Gaussian missing mass distribution of
width $\sigma$, about 87\% of the signal is contained in a bin
$\Delta M_{\rm missing}=3\sigma$, that is $M_{\rm missing}=M_H \pm 1.5\sigma$.

We could consider Higgs production in other diffractive channels, such as diffractive
production accompanied by
proton dissociation ($pp \to M_1+H+M_2$), or central inelastic production
($pp \to p+(M \to HX)+p$)  \cite{INC}.  However, we do not gain
much as compared to  the usual totally
inclusive production --
there is no precise missing mass measurement, no selection rule to suppress the $\bb$
background and more serious pile-up problems.  The somewhat smaller density of soft secondary hadrons in the
Higgs rapidity region does not compensate for the much smaller statistics
(cross sections) in diffractive processes.

\section{Exclusive SUSY  $H\ra\bb$  signals}

\begin{figure}
\begin{center}
\centerline{\epsfxsize=0.5\textwidth\epsfbox{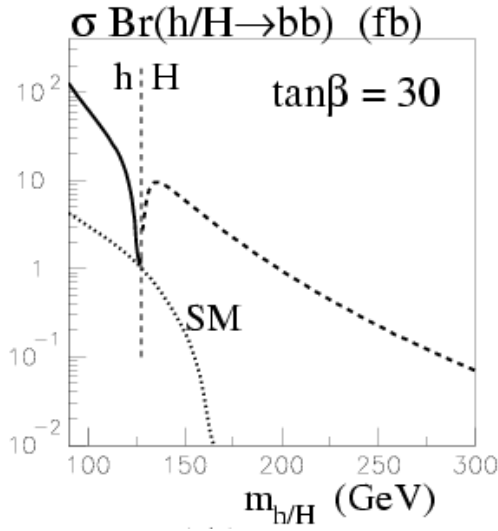}}
\caption{The cross section times the $b \bar b$ branching ratio for central exclusive production of $h$ and $H$ MSSM Higgs bosons at the LHC for
$\tan\beta=30$. Also shown (by the dotted curve) is the cross section times the branching ratio for SM Higgs production.}  
\label{fig:hH}
\end{center}
\end{figure}

To be specific, we discuss the three neutral Higgs bosons of the MSSM
model: $h,H$ with CP=1 and $A$ with CP=--1 \cite{CH}.  There are regions of MSSM
parameter space, for example, the intense coupling regime \cite{bdn}, 
 where the conventional signals ($\gamma\gamma, WW, ZZ$
decays) are suppressed, but where the exclusive subprocess $gg\ra H \ra \bb$ is strongly
enhanced \cite{KKMR04}.  This is evident from Fig.~$\ref{fig:hH}$, which shows the cross section for CEP production of $h,H$ bosons as function of their mass for
$\tan\beta=30$.
Here, and in what follows, we use version 3.0 of the HDECAY code \cite{HDEC}.

Taking, for example, for $M_A$ = 130 GeV and tan$\beta$ = 50, we have
$M_h$ = 124.4 GeV with $S/B=71/9$ events, $M_H$ = 135.5 GeV with $S/B=124/6$
events and $M_A$ = 130 GeV with $S/B=0.17$, so both
$h$ and $H$ should be clearly visible. (Again, for reference,
we assume that $\Delta M_{\rm missing}=3$ GeV can be achieved.) 
Let us emphasize that the  intense coupling 
regime of the MSSM \cite{bdn} is especially forward proton  friendly,  
and in this particular case the tagged-proton approach may well
be {\it the} discovery channel.
 
 The decoupling regime
($M_A \gtrsim 2M_Z$ and tan$\beta\gtrsim 5$) is another example where the exclusive
signal is of great value.   In this case $h$ is indistinguishable from a SM Higgs, and so the discovery
of $H$ is crucial to establish the underlying dynamics. 
 
If the exclusive cross sections for scalar and pseudoscalar Higgs production were comparable,
it would be possible to separate them readily by a missing mass scan, and by the study of
azimuthal correlations between the outgoing protons.  Unfortunately,
pseudoscalar exclusive production is strongly suppressed by the P-even selection.
Maybe the best chance to identify the $A(0^-)$ boson is through the double-diffractive dissociation
process, $pp\ra X+A+Y$, where both
protons dissociate \cite{KKMR04}.

\section{Detecting the Higgs in the $WW$ channel}            

The analysis in the previous sections was focused primarily
 on light SM and MSSM Higgs production, with the Higgs 
decaying to 2 $b-$jets. The potentially copious $b-$jet (QCD) background is controlled by
 a combination of the spin-parity 
selection rules
and the mass resolution from the forward proton 
detectors. The missing-mass resolution is especially critical in controlling the background,
since poor resolution would allow more background events into the mass window around the resonance.

Whilst the $b \bar b$ channel is very attractive theoretically,
allowing direct access to the dominant decay mode
of the light Higgs boson, there are some basic 
problems which render it challenging from an experimental perspective,
see \cite{cox2} for details.
First, it relies heavily on the quality of
 the mass resolution from the proton taggers to suppress 
the background. 
Secondly, triggering on the relatively low-mass dijet signature of the
$H \rightarrow b \bar b$ events is a challenge
for the Level 1 triggers of both ATLAS and CMS. And, thirdly, this
measurement requires double $b-$tagging, with a corresponding price to pay for the
tagging efficiency.

In \cite{krs,cox2}, attention was turned to
 the $WW^*$ decay mode of the light Higgs, and for a Higgs mass above 
 the $WW$ threshold, to the $WW$ decay mode. This channel 
does not suffer from any of the above problems: suppression 
of the dominant backgrounds does not rely so strongly on the mass resolution of the 
detectors, and, certainly, in the semi-leptonic decay channel of the $WW$ system, 
the Level 1 triggering is not a problem. The advantages of forward proton tagging are, however, 
still explicit. Even for the double leptonic decay channel (i.e. with two leptons and two
final state neutrinos), the mass resolution will be very good, and, of course, 
observation of the Higgs in the double-tagged channel immediately 
establishes its quantum numbers. It is worth mentioning
that the mass resolution should improve with increasing Higgs mass.  
Moreover, the semileptonic `trigger cocktail' may allow
a combination of signals, not only from $H\to WW$ decays, but also from
the $\tau\tau$, $ZZ$ and even the semileptonic $b-$decay channels.

\begin{figure}
\begin{center}
\centerline{\epsfxsize=0.8\textwidth\epsfbox{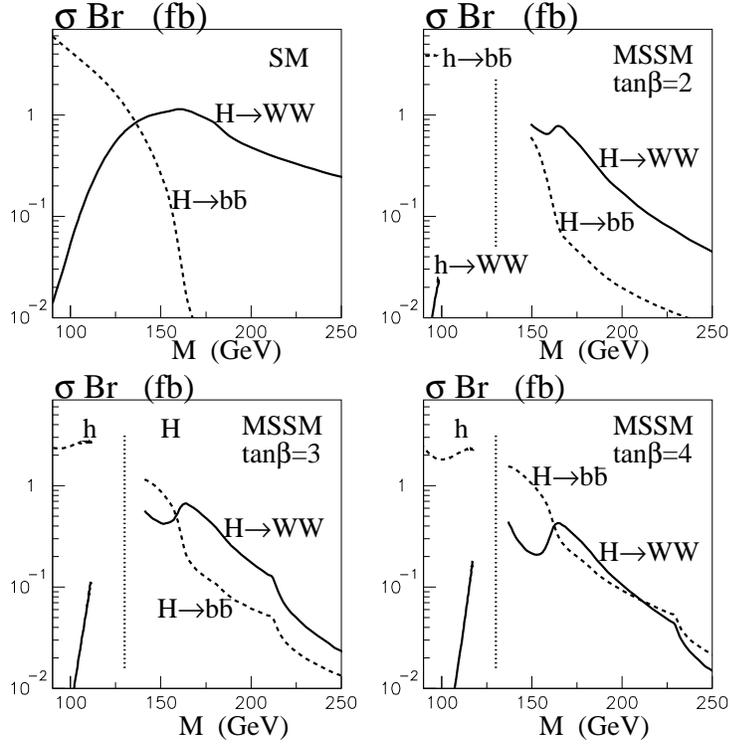}}
\caption{The cross section times branching ratio 
for the central exclusive production of the MSSM Higgs boson (for three values of
$\tan\beta = 2,3,4$),
as a function of the Higgs mass, in the $WW$ and $b \bar b$ decay channels. The cross section
for the CEP production of a SM Higgs boson is also shown.}  
\label{fig:tanbeta}
\end{center}
\end{figure}

In Fig.~\ref{fig:tanbeta} we show the cross section for 
the process $pp \rightarrow p+H+p \rightarrow p+WW+p$ 
as a function of the Higgs mass $M_H$ at the LHC. The increasing 
branching ratio 
to $WW^{(*)}$ 
as $M_H$ increases (see for example \cite{CH}) 
compensates for the falling central exclusive production cross section. 
For comparison, we also show the cross 
section times branching ratio for $pp \rightarrow p+H+p \rightarrow p+b \bar b +p$.
For reference purposes, the cross sections in Fig.~\ref{fig:tanbeta}  are normalized in such a way
that $\sigma_H = 3$~fb for $M_H = 120$~GeV. 
In  Fig.~\ref{fig:tanbeta} we show also the results for $\tan\beta=2,3,4$. Evidently the
expected CEP yield is also promising in the low  $\tan\beta$ region.

Experimentally, events with two $W$ bosons in the final
 state fall into 3 broad categories  --- fully-hadronic,
semi-leptonic and fully-leptonic --- depending on 
the decay modes of the $W$'s. Events in which at 
least one of the $W$s decays in either the electron or muon channel 
are by far the simplest, and Ref.~\cite{cox2} focuses mainly
on these semi- and fully-leptonic modes. 
As mentioned above, one of the attractive features of the $WW$ channel 
is the absence of a relatively large 
irreducible  background, in contrast to the large 
central exclusive $b \bar b$ QCD background  in the case of $H \rightarrow b \bar b$,
suppression of which
relies strongly on the experimental missing-mass resolution 
and di-jet identification.
The primary  exclusive  backgrounds for the $WW$ channel
 can be divided into two broad categories: 
\begin{enumerate}
\item central production of a $WW^*$ pair $pp\to
p+(WW^*)+p$ from either the (a) $\gamma\gamma\to WW^*$ or (b)
$gg^{PP}\to WW^*$ subprocess,
\item 
the $W$-strahlung process  $pp\to p+Wjj+p$
originating from the $gg^{PP}\to Wq \bar q$
subprocess, where  the $W^*$ is `faked' by the
two quarks.
\end{enumerate}
As shown in \cite{cox2}, over a wide region of  Higgs masses
the photon-photon initiated backgrounds are strongly suppressed
if we require that the final leptons and jets are central
and impose cuts on the transverse momenta of the protons in the taggers.
Moreover, our estimates show that the QCD quark-box-diagram contribution
from the $gg^{PP}\to WW^*$ subprocess
is very small.

The most important  background, therefore, comes from the second category, i.e. from
the $W$-strahlung process. Here
we have to take into account the $J_z=0$ projection of this amplitude, which requires
a calculation of the individual helicity amplitudes.  This was done in \cite{krs}
using the spinor technique of Ref.~\cite{kwjs}.
The analysis in \cite {krs,cox2}
 shows that this background can be manageable with carefully choosen experimental
cuts.
For $M_H = 140$~GeV we expect 19 exclusive  $H\to WW$ events
for an LHC luminosity of 30~fb$^{-1}$.
Note that the largest loss of events in the $WW$ case is caused by
the Level 1 trigger efficiency, and we expect significant improvements 
here.

\section{Related processes: checks of the predicted exclusive Higgs yield}

As discussed above, the exclusive Higgs signal is particularly clean, and the signal-to-background
ratio is  favourable.
 However, the expected number of events in the SM case is low.
Therefore it is important to check the predictions for exclusive Higgs production
by studying processes mediated by the same mechanism, but
with rates which are sufficiently high, so that they may be observed at the Tevatron
(as well as at the LHC).  The most obvious examples are those in which the Higgs 
is replaced by either a dijet system, a $\chi_c$ or $\chi_b$ meson, or
by a $\gamma \gamma$ pair, see Fig.~$\ref{fig:H}$. 

First, we discuss the exclusive production
of a pair of high $E_T$ jets, $p\bar {p} \to p+jj+\bar {p}$ \cite{KMR,INC}.
This would provide an effective $gg^{PP}$ `luminosity
monitor' just in the kinematical region of the Higgs
production. The corresponding cross section was evaluated to
be about 10$^4$ times larger than that for the SM Higgs boson.
Thus, in principle,
this  process appears to be an ideal `standard candle'.  The expected cross section is rather large,
and we can study its behaviour as a function of the mass of the dijet
system.  This process is being studied by the CDF collaboration. Unfortunately, in the present CDF environment, the separation
of exclusive events is not unambiguous.
At first sight, we might expect that the exclusive dijets form a narrow peak,
sitting well above the background, in the
distribution of the ratio
\begin{equation}
R_{jj}=M_{{\rm dijet}}/M_{\rm {PP}}
\end{equation}
at $R_{jj}=1$, where $M_{\rm {PP}}$ is the invariant energy of the incoming
 Pomeron-Pomeron system.  In reality
the peak is smeared out due to hadronization and the jet-searching algorithm.
Moreover, since $M_{{\rm dijet}}$ is obtained from measuring just the two-jet part of the exclusive signal, there will be a `radiative tail' extending to lower values of $R_{jj}$.

\begin{figure}
\begin{center}
\centerline{\epsfxsize=0.9\textwidth\epsfbox{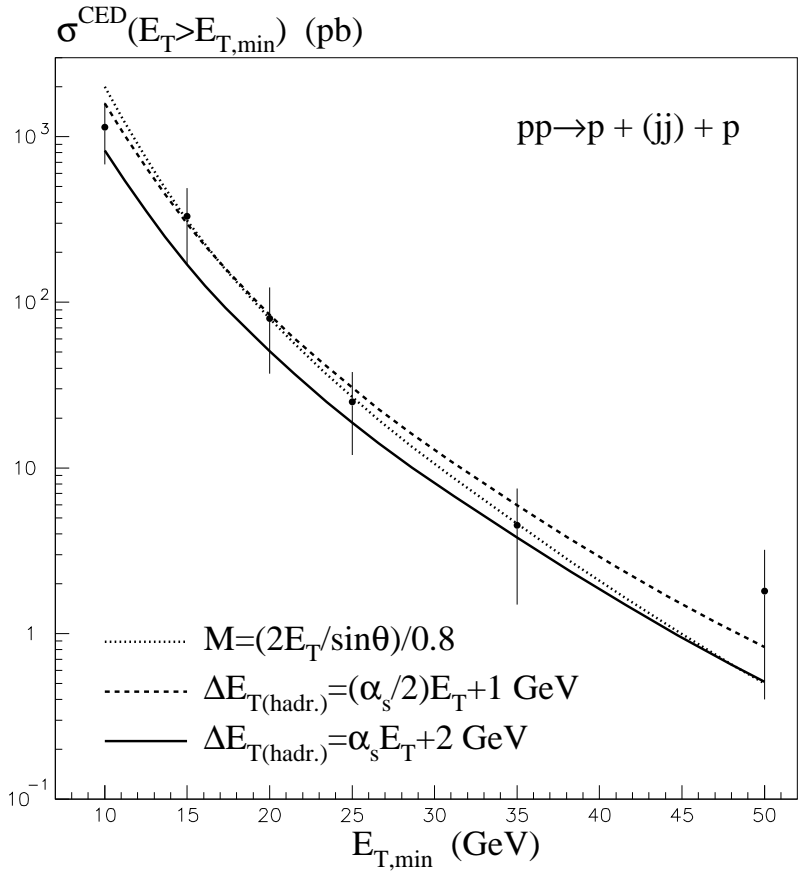}}
\caption{The cross section limit for `exclusive' dijet production at the Tevatron as a function $E_{T,{\rm min}}$ as measured by CDF \cite{CDFchi}.  These preliminary CDF data correspond to the cross section integrated over
the domain $R_{jj} =M_{\rm dijet}/M_{\rm PP} > 0.8$ and $E_T > E_{T,{\rm min}}$.
The curves are the pure exclusive cross section
calculated \cite{INC} using the CDF event selection. Different hadronization corrections were applied.
The solid curve is obtained assuming that, after the hadronization, the measured jet
transverse energy $E_T$ is less than the parton (gluon) transverse
energy by $\Delta E_T = E_{T,{\rm gluon}}-E_{T} =
\alpha_s(E_T) E_T + 2$ GeV; while for the dashed
 curve it is assumed that $\Delta E_T$ is halved, i.e.
$\Delta E_T =(\alpha_s(E_T)/2) E_T + 1$ GeV. The dotted curve
is calculated assuming $E_{T,{\rm gluon}}=E_T$, but with the mass of the
whole central system (which determines the incoming
gluon-gluon luminosity) enlarged according to the $R_{jj}$
ratio --  $M_{\rm PP}
=(2E_T/{\rm sin}\theta)/0.8$, where $\theta$ is the jet polar angle in the
dijet rest frame.}
\label{fig:JJ}
\end{center}
\end{figure}

For jets with $E_T=10$ GeV and a jet cone $R<0.7$, more than 1 GeV will be lost
outside the cone, leading to (i) a decrease of the measured jet energy of about 1-2 GeV,
and, (ii) a rather wide peak ($\Delta R_{jj}\sim \pm 0.1$ or more) in the $R_{jj}$ distribution.
The estimates based on Ref.~\cite{INC} (see also \cite{KMRSgam}) give an exclusive cross section for dijet
production with $E_T> 10,  25,  35, 50$ GeV, with values which are rather close
to the recent CDF limits \cite{CDFchi}.  The comparison is shown in Fig.~$\ref{fig:JJ}$. In particular, for $E_T> 50$ GeV, we predict an exclusive cross section of about 1 pb \cite{INC}, which agrees well with the current
CDF upper limit obtained from events with $R_{jj} > 0.8$. 
As discussed above,
one should not expect a clearly `visible' peak in the CDF data for $R_{jj}$ close to 1.
 It is worth mentioning that the CDF measurements have already started to reach values of the invariant mass of the Pomeron-Pomeron system in the SM Higgs mass range. 

\begin{figure}
\begin{center}
\centerline{\epsfxsize=0.8\textwidth\epsfbox{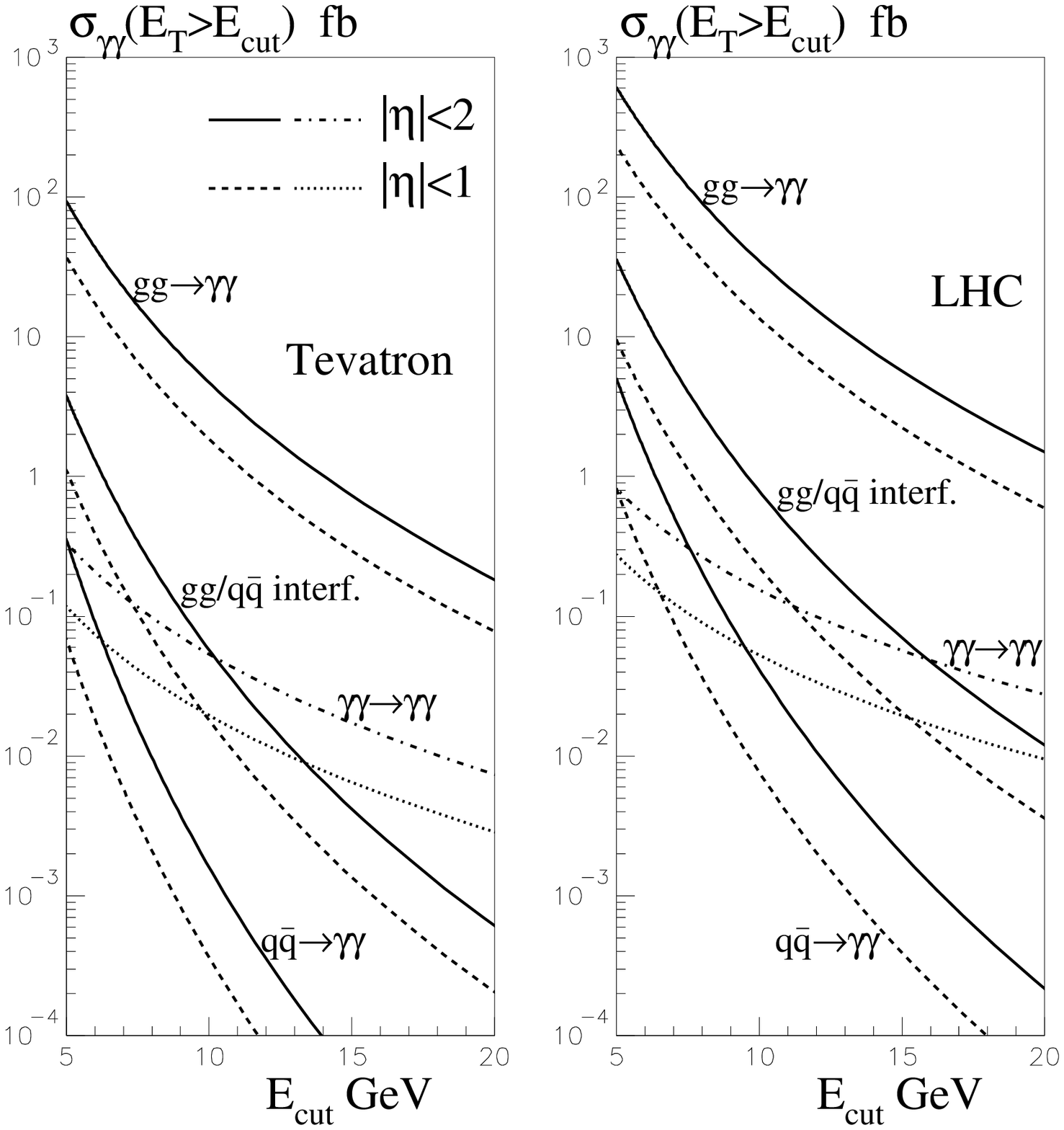}}
\caption{The contributions to the cross section for exclusive $\gamma \gamma$ production
from $gg$ and $\qq$ exchange at the Tevatron and the LHC.   Also shown is the
contribution from the QED subprocess $\gamma \gamma \to \gamma \gamma$.  For each component we
show the cross section restricting the emitted photons to have $E_T>E_{\rm cut}$
and to lie in the centre-of-mass
rapidity interval $|\eta_{\gamma}|<1$ (or $|\eta_{\gamma}|<2$).
The figure is taken from Ref.~\cite{KMRSgam}.}
\label{fig:gamma}
\end{center}
\end{figure}

An alternative `standard candle' process is exclusive double-diffractive $\gamma\gamma$ production
with high $E_T$ photons, that is $p\bar {p} \to p+\gamma \gamma +\bar {p}$ \cite{INC,KMRSgam}.
Here there are no problems with hadronization or with the identification of the jets.
On the other hand, the exclusive cross section is rather small. 
The predictions of the cross section for
 exclusive $\gamma \gamma$ production are shown in Fig.~$\ref{fig:gamma}$.

The CDF collaboration has reported \cite{CDFchi}
a preliminary result for exclusive $\chi_c$ production.
Although it is consistent with perturbative QCD expectations \cite{KMRSchi},
the mass of the $\chi_c$-boson, which drives the scale of the process, is too
low to justify just the use of perturbative QCD \footnote {Even lower scales
correspond to the fixed target central double diffractive meson resonance production
observed by the WA102 collaboration at CERN\cite{WA102}}. 
 Therefore,
it is intriguing that the qualitative features of the observed $p_t$ and
azimuthal angular distributions appear to be in good agreement with the
perturbatively based expectations\cite{KMRtag}. However,
in Ref.~\cite{KMRSchi}, it was found that both a Regge formalism and perturbative QCD predict essentially the
same qualitative behaviour for the central double-diffractive production of
`heavy' $\chi_c(0^{++})$  and $\chi_b(0^{++})$ mesons\footnote{Note that the 
results for  $\chi_c(0^{++})$, given in \cite{KMRSchi}, should be
decreased by a factor of 1.5 due to the new value of the total
$\chi_c(0^{++})$ width in PDG-2004 \cite{pdg}.  Thus, the predicted cross section for $\chi_c \to J/\psi + \gamma \to \mu \mu \gamma$ is now about 300 pb; with the CDF experimental cuts, it becomes about 50 pb.}.
Due to the
low scale, $M_\chi /2$, there is a relatively small contribution coming from the process,
in which the incoming protons dissociate. Therefore simply
selecting events with a rapidity gap on either side of the $\chi$, almost
ensures that they will come from the exclusive reaction, $p\bar {p} \to p\ +\ \chi\ +\ \bar {p}$.
Although exclusive $\chi$ production is expected to dominate, the  predicted\cite{KMRSchi} event rates are large
enough to select double-diffractive dissociation events with large transverse
energy flows in the proton fragmentation regions. Such events are particularly
interesting.  First, in this case, the large value of
$E_T$ provides the scale to justify the validity, and the reasonable
accuracy, of the perturbative QCD calculation of the cross section. Next, by measuring the
azimuthal distribution between the two $E_T$ flows, the
parity of the centrally produced system can be determined.

Another possible probe of the exclusive double-diffractive
formalism would be to observe central open $\bb$ production;
namely $b,{\bar b}$ jets with $p_t\gtrsim m_b$.
Again, this would put the application of perturbative QCD on a sounder footing.
It would allow a check of the perturbative formalism, as well as
a study of the dynamics of $\bb$ production.

\section{Conclusion}

The installation of proton-tagging detectors in the forward region around 
ATLAS and/or CMS would add unique capabilities 
to the existing LHC experimental programme. The current calculations of the rates of CEP
processes show that 
 there is a real chance that new heavy particle production could
 be observed in this channel. For the 
Standard Model Higgs, this would amount to a direct determination of its quantum numbers, 
with an integrated luminosity of order 
30 fb$^{-1}$. For certain MSSM scenarios, the tagged-proton channel may even be the discovery channel.
 At higher luminosities, proton tagging 
may provide direct evidence of CP-violation within the Higgs sector. There is also a rich
QCD, electroweak, and 
more exotic, physics menu. This includes searches for extra dimensions, 
gluino and 
squark production, gluinonia, and, indeed, any object which has $0^{++}$ or $2^{++}$ quantum 
numbers and couples strongly to gluons \cite{INC}.
 
Here we focused on the unique advantages of CEP
Higgs production. The missing mass, $M_{\rm missing}$, measured by
the forward proton detectors can then be matched with the mass $M_{\bb}$ from the main decay
mode, $H \to \bb$.  Moreover the QCD $\bb$ background is suppressed by a $J_z=0$ selection rule.
The events are clean, but the predicted yield is low: about 10 events,
after cuts and acceptance, for an integrated luminosity of
${\cal L}=30~{\rm fb}^{-1}$.  The signal-to-background ratio is about 1,
depending crucially on the accuracy with which $M_{\rm missing}$ can be measured.

From the experimental perspective, the simplest channel to observe a Higgs Boson of mass between
140 GeV and 200 GeV
is the $WW$ decay mode. According to studies in \cite {cox2}, there will be a detectable signal, and the backgrounds
should be controllable. 
 
We have emphasized the importance of checking the perturbative QCD predictions by
observing analogous CEP processes, with larger cross sections, at the
Tevatron.

\section*{Acknowledgements}

This work was supported by
the UK PPARC, by a Royal Society FSU grant, by grants INTAS 00-00366,
RFBR 04-02-16073 and 01-02-17383 and by the Federal Program of
the Russian Ministry of Industry, Science and Technology SS-1124.2003.2.
ADM thanks the Leverhulme Trust for an Emeritus Fellowship.

\end{document}